\documentclass[aps, twocolumn]{revtex4-1}

\pdfoutput=1

\usepackage{amsfonts}
\usepackage{amsmath}

\usepackage{hyperref}
\usepackage{bm}
\usepackage{graphicx}
\usepackage{threeparttable}  

\usepackage[pdftex]{color}
\definecolor{spring}{rgb}{0.7,0.9,0.7}
\definecolor{brick}{rgb}{0.7,0.2,0.1}
\definecolor{redHL}{rgb}{1.0,0.5,0.5}

\def\qeff{Q_{\rm eff}}
\begin{document}

\title{Resonant Dampers for Parametric Instabilities in Gravitational Wave Detectors}

\author{S. Gras}
\email[E-mail me at: ]{sgras@ligo.mit.edu}
\author{P. Fritschel}
\author{L. Barsotti}
\author{M. Evans}
\affiliation{Massachusetts Institute of Technology, 185 Albany St. NW22-295, 02139 MA, USA}
\pacs{???}
\date{today}

\begin{abstract}
Advanced gravitational wave interferometric detectors will operate at their design sensitivity with nearly $\sim$1MW of laser power stored in the arm cavities. Such large power may lead to the uncontrolled growth of acoustic modes in the test masses due to the transfer of optical energy to the mechanical modes of the arm cavity mirrors. These parametric instabilities have the potential of significantly compromising the detector performance and control. 
Here we present the design of ``acoustic mode dampers" that use the piezoelectric effect to reduce the coupling of optical to mechanical energy. Experimental measurements carried on an Advanced LIGO-like test mass shown a 10-fold reduction in the amplitude of several mechanical modes, thus suggesting that this technique can greatly mitigate the impact of parametric instabilities in advanced detectors. \end{abstract}

\maketitle

\section{Introduction}
The network of advanced gravitational wave detectors currently under construction (two LIGO \cite{0264-9381-27-8-084006} detectors in the USA, the Advanced Virgo \cite{AdvancedVIRGO} detector in Italy, and Kagra \cite{0264-9381-29-12-124007} in Japan) promises to open the new window of gravitational wave astronomy within this decade.

Sensitivity to gravitational wave strains of order $10^{-24}$ requires high optical power circulating in the arm cavities of these detectors. For instance, up to 750 kW of optical power will be sustained in the steady-state regime inside the Advanced LIGO arm cavities.

It has been experimentally observed that the stored energy in a resonant cavity can leak from the optical modes to the mechanical modes of the cavity mirrors via a 3-mode interaction \cite{PhysRevA.78.023807}. Given sufficiently high circulating optical power, and mirror materials with very low mechanical loss as required to avoid thermal noise, the uncontrolled growth of test mass acoustic modes can lead to Parametric Instabilities (PI) \cite{Evans2010665, 0264-9381-27-20-205019}.
If left unaddressed PI will prevent high power operation,
 and thus limit the astrophysical output of gravitational wave detectors.
While Advanced LIGO will serve as the primary example in this paper,
 all advanced gravitational wave detectors are susceptible to these instabilities.

Several schemes have been proposed to damp PI~\cite{PhysRevLett.94.121102, PhysRevA.81.013822}. 
In particular, solutions directly applicable to Advanced LIGO involve active damping of acoustic modes by means of the test mass electro-static drive actuators~\cite{Miller2011788}, and thermal tuning of the optical modes using the test mass ring heaters~\cite{CQG_0264-9381-26-13-135012}. A significant constraint on any technique is that it must preserve the inherently low mechanical loss of the test mass in the gravitational wave frequency band to maintain a low level of thermal noise.

Here we present a novel method to passively control PI by
 reducing the Q-factor of the test mass acoustic modes with
 small resonant dampers.
These ``acoustic mode dampers"  (AMD) dissipate the strain energy of the test mass
 mode through a resistive element after converting it to electrical energy
 via the piezoelectric effect (see figure \ref{fig:spring-mass}).
 
The resonant nature of the AMD allows it to
 effectively damp test mass acoustic modes without introducing significant mechanical
 loss at lower frequencies where thermal noise can limit detector performance.
With respect to other proposed solutions,
 this approach has the advantage of being simple, self-contained, and completely passive.
Models indicate that AMDs can provide a broadband
 reduction in the Q of mechanical modes relevant to PI,
 and are therefore particularly beneficial in the presence of a large number of unstable modes.

The structure of this paper is as follows.
In section \ref{sec:PI} we set the stage by giving a brief overview of parametric instabilities,
 including equations of particular relevance to evaluating AMD performance. 
Section \ref{sec:ds} presents a simple 1-dimensional model of the AMD which highlights
 the principles of AMD operation. 
This is followed by a description of the detailed finite element model (FEM) used
 to analyze the AMDs ability to suppress PI when attached to an Advanced LIGO test mass.
The FEM predictions are compared with experimental results from a full-scale prototype
 in section \ref{sec:er}.
Finally, in section \ref{sec:tn} we discuss an AMD design that
 will provide Advanced LIGO with
 protection from instability, without significantly increasing test mass thermal noise.

\section{Parametric Instabilities (PI)}\label{sec:PI}
\begin{figure}[ht!]
      \center{\includegraphics[width=0.5\textwidth]{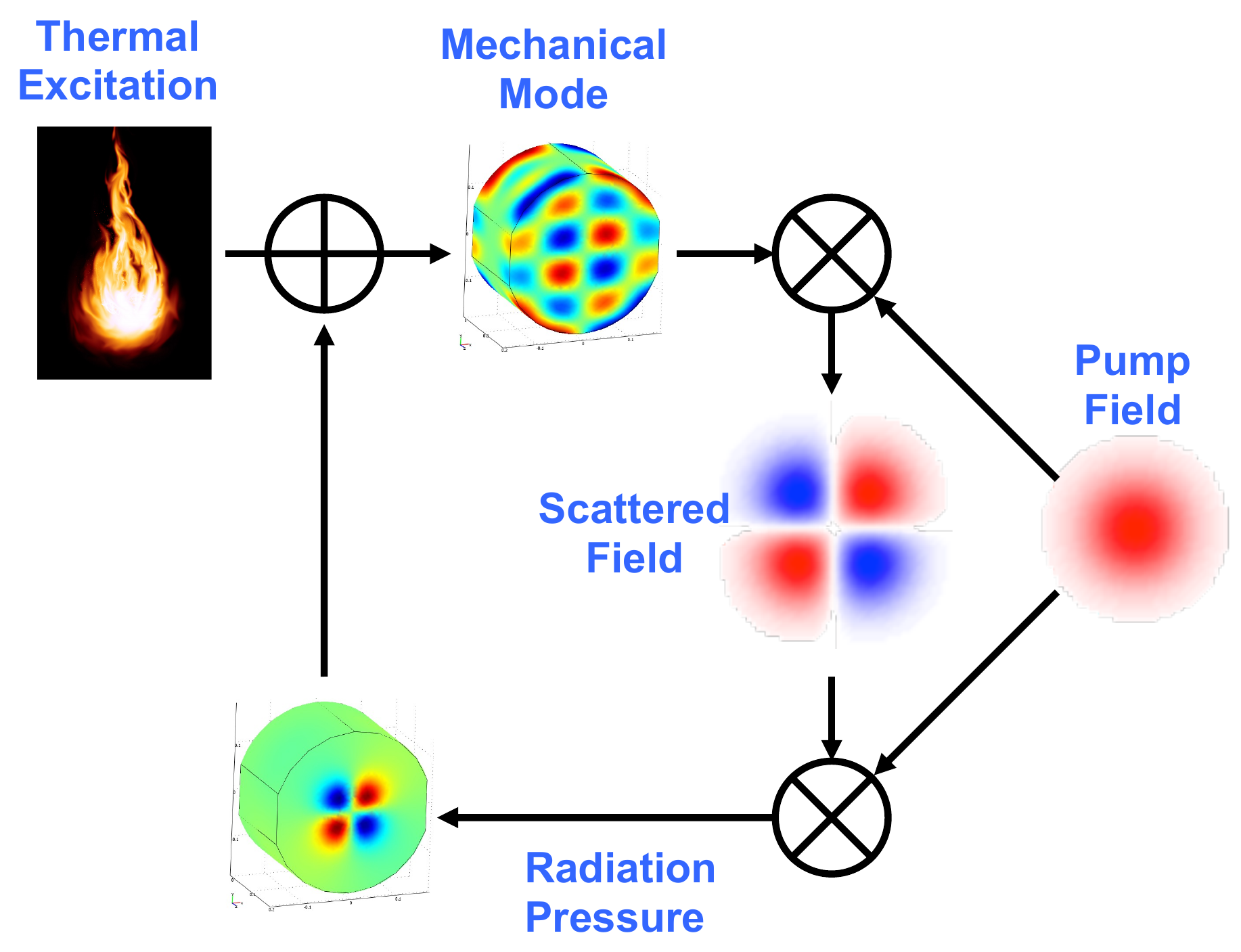}
       \caption{\label{fig:ShearPZT} PI described as a positive feedback process.
       A steady state cavity field inside the interferometer arm cavity is distorted by 
       scattering off a vibrating mirror surface.
       Two transverse optical sidebands are created.
       Both sidebands exert force on the test mass via radiation pressure.
       When the energy dissipation of the acoustic mode and the rate of 
       work done by the radiation pressure is unbalanced,
       one of the sidebands excites the exponential growth of the acoustic mode amplitude.
       The dynamic of this process is commonly described in terms of the parametric
       gain $R$, with $R>1$ in the case of instability.}}
\end{figure}

The acousto-optic interactions responsible for parametric instabilities have been extensively studied~\cite{Braginsky2001331,PhysRevLett.94.121102,Evans2010665}.
They consist of a scattering process and radiation pressure operating together
 in an optical cavity in a closed-loop manner.
 The graphical representation of this process is shown in Fig.\ref{fig:ShearPZT}.

The e-folding growth time, or ``ring-up time'',
 of an acoustic mode in the presence of a 3-mode interaction is
 $\tau =2Q_{m}/(\omega_{m}(R-1))$ \cite{Ju2006360}, where $Q_m$ and $\omega_m$ are the Q-factor and angular frequency of the mode, respectively, and $R$ is the parametric gain.
When $R > 1$ the amplitude increases exponentially until a saturation point is reached \cite{arXiv:1303.4561v2}.
The parametric gain $R$ can be defined as
\begin{equation}
 R = 4 \pi^2 P_{c}\Lambda Q_{m}\times\frac{\nu_{o}}{\delta\nu_{hom}}\times\frac{\nu_{o}}  {\delta\nu_{00}}\times\Gamma(\Delta\omega)  \label{eqn:R}
\end{equation}
where $P_{c}$ is the optical power circulating in the arm cavity, $\nu_{o}$ is the optical frequency of the light, and $\delta\nu_{00}$ and $\delta\nu_{hom}$ are the cavity linewidths (full width, half maximum) for the fundamental and the higher-order optical mode, respectively. The parameter $\Lambda$ measures the spatial overlap between the acoustic and the higher-order optical modes; $\Gamma$ is representative of the interferometer optical configuration and is a function of the 3-mode interaction tuning 
$\Delta\omega = \omega_{m} - 2\pi \Delta\nu_{hom}$, where $\Delta\nu_{hom}$ is the frequency difference between the fundamental and higher-order optical modes.
For $\Delta\omega\rightarrow 0$, the parameter $\Gamma$ reaches its
 maximum (see \cite{Strigin200710} for a more detailed description).

Unstable acoustic modes with parametric gain up to $R \simeq 100$ may arise in Advanced LIGO
 in the  10-90 kHz band~\cite{Evans2010665}. To prevent these instabilities, a damping mechanism must be introduced
 to reduce the Q-factor of all unstable acoustic modes in this frequency band
  {\it without introducing excess thermal fluctuation in the detection band} of 10~Hz to 1~kHz.

\section{Model of the Acoustic Mode Damper (AMD)} \label{sec:ds}

In order to reduce the Q test mass acoustic modes we designed a resonant AMD which can be attached to the test mass and provide dissipation via the piezo-electric effect. 
\begin{figure*}[ht!]
      \center{\includegraphics[width=0.8\textwidth]{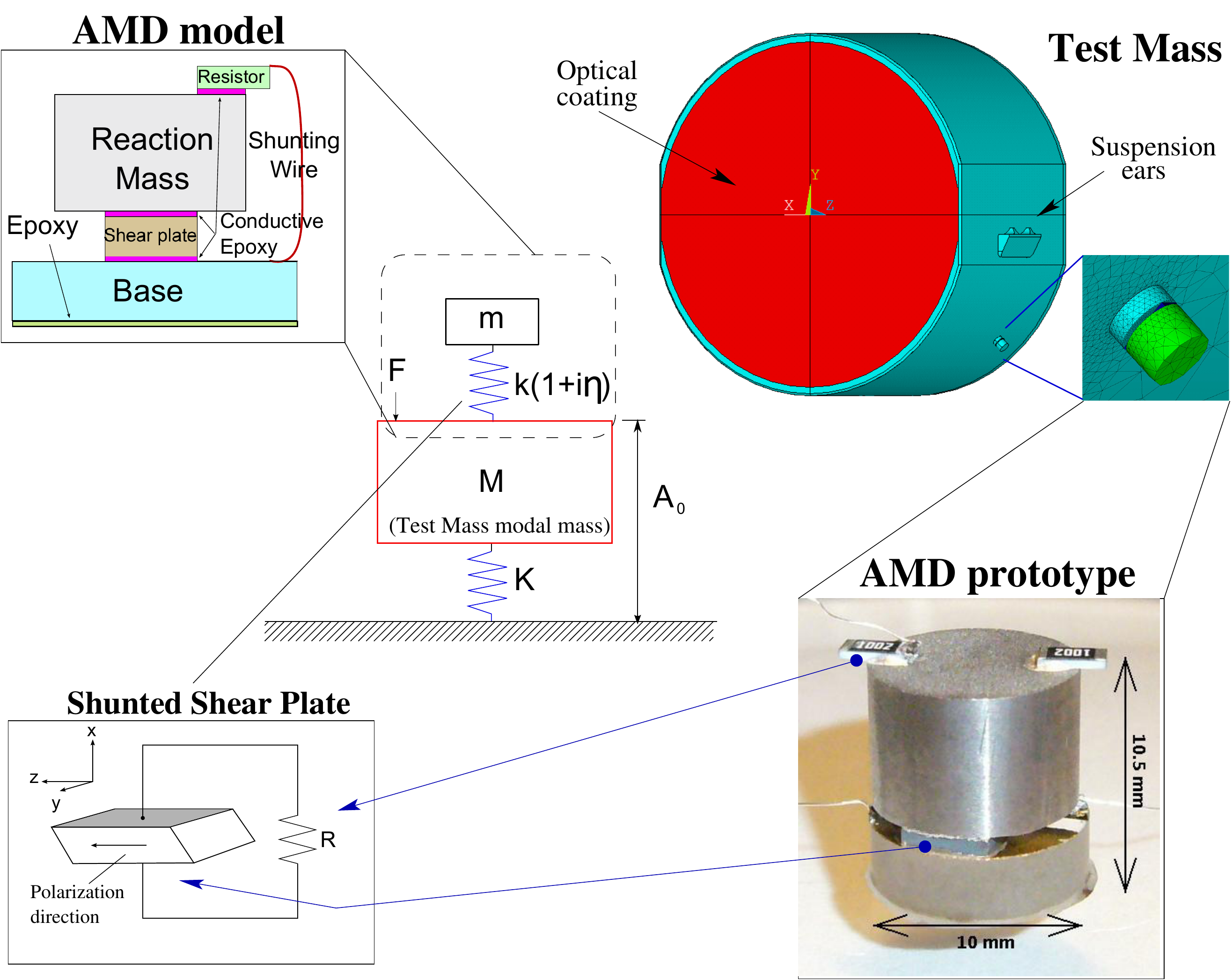}}
       \caption{\label{fig:spring-mass} Overview of the Acoustic Mode Damper (AMD). The AMD can be described as a small spring-mass system attached to a larger mass $M$. $M$ represents a modal mass of an acoustic test mass mode which vibrates due to an external force $F$ (radiation pressure). Such induced vibration causes deformation of the lossy spring of the damper. The non-zero loss angle of the damping spring (piezoelectric material) results in the lag of strain with respect to stress and thus energy dissipation. As a consequence of this dissipation process the amplitude of the acoustic mode vibration is reduced.}
\end{figure*}

In this section we first describe the interaction between the AMD and the test mass with a simple 1-D model, then we present a complete Finite Element Model of the entire system.

\subsection{Simplified 1-D Model}\label{sec:1D}

The AMD and test-mass system can be described as a pair of coupled oscillators with a large mass ratio.
The AMD mass $m$ is attached to the much more massive test mass via piezo electric shear plates,
 which are modeled as a lossy spring with complex spring constant of magnitude $k$ and loss angle $\eta$.

The test mass acoustic mode for which we would like to estimate the impact of the AMD
 is simplified in this model to a mass $M$, equal to the modal mass of the acoustic mode,
 attached to a fixed reference by a lossless spring $K$.
The coupled systems is then excited by the radiation pressure force $F$
 applied to the TM mode, as shown in Fig.~\ref{fig:spring-mass}.

\begin{figure}[h!]
     \center{\includegraphics[width=0.5\textwidth]{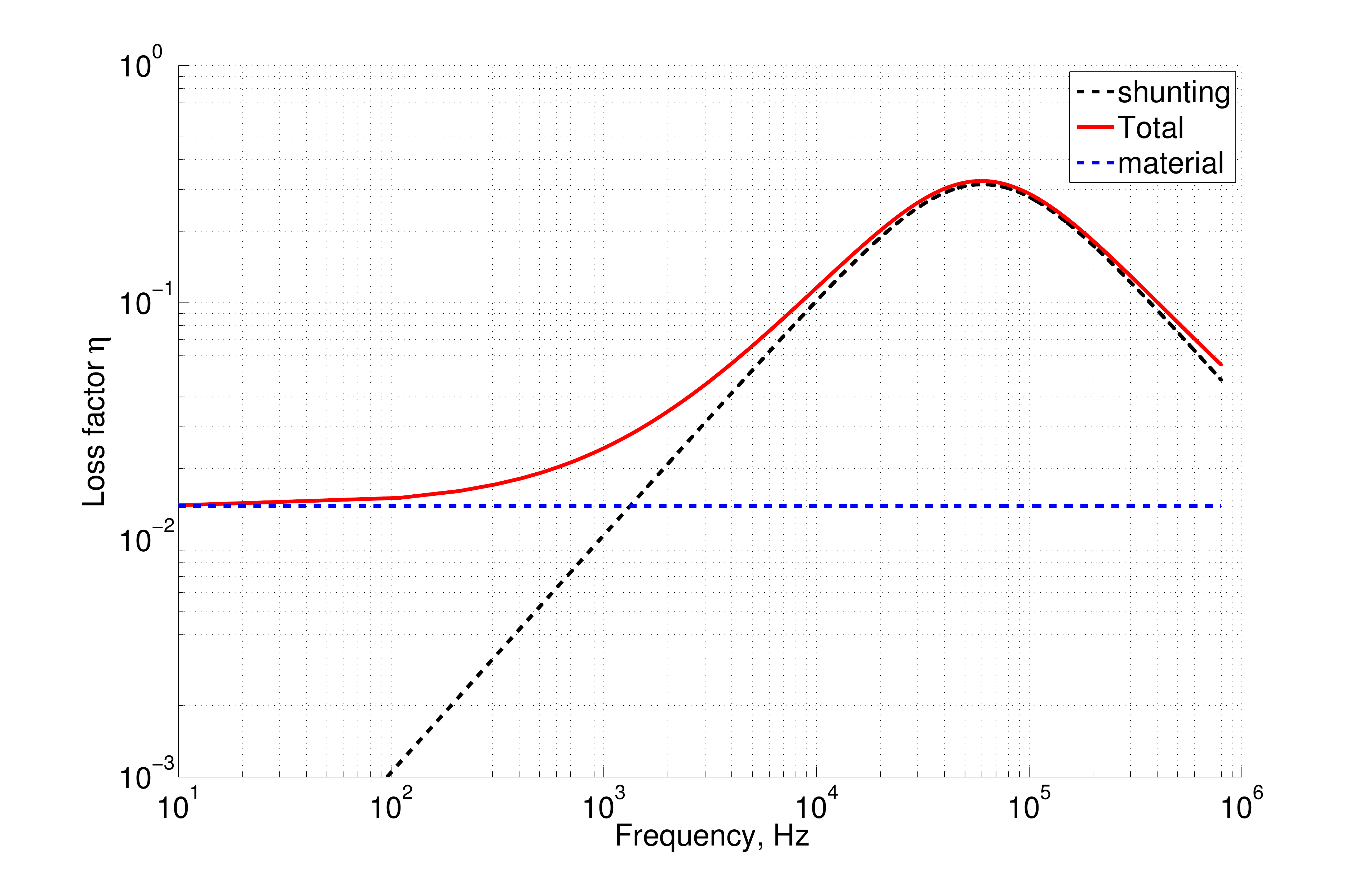}}
       \caption{\label{fig:PZTloss} A viscous-like behavior of the loss angle for shunted piezoelectric material. The total loss factor is a combination of the shunting loss and structural loss of the piezoelectric material. The maximum peak position is located at ~40 kHz which corresponds to the shunt with 10 k$\Omega$ resistor. The larger the resistor the lower is frequency of the peak loss. The peak height is proportional to electromechanical coupling coefficient $k_{15}$. Note, the loss angle at low frequency section is mainly dominated by the structural loss angle of piezoelectric material.}
     \end{figure}

At frequencies near the resonance of the AMD,
 the lossy spring produced by the piezoelectric material and resistive load
 will dissipate the energy of the excited acoustic mode, as seen in Fig.~\ref{fig:PZTloss}.
 
For this system of coupled oscillators, the amplitude $A(\omega)$ of the acoustic mode driven by force
 $F$ at angular frequency $\omega$ is
\begin{gather}
A(\omega) = \frac{F}{M \omega^2} \sqrt{
 \frac{\epsilon^2 + \eta^2} {(\delta \epsilon + \mu)^2 + \eta^2 (\delta + \mu)^2 }} \\
 \mbox{where}~~ \delta = 1 - \omega_0^2 / \omega^2, ~
   \epsilon = 1 - \omega^2 / \omega_D^2   \label{eq:ampl}  \nonumber \\
 \quad \omega_0^2 = \frac{K}{M}, ~ \omega_D^2 = \frac{k}{m} ,  ~ 
 \mbox{and} ~ \mu = \frac{m}{M}. \nonumber \\
\end{gather}

The resulting effective Q-factor is
\begin{equation}
\qeff= \frac{\max ( A(\omega) )}{ A(\omega=0)} \simeq \frac{\eta^2 + (1 - \rho)^2}{\eta \mu \rho}
\label{eq:qeff}
\end{equation}
where  $\rho = \omega_{0} / \omega_{1}$, and we assume $\mu \ll 1$.

When the acoustic mode resonance is near that of the AMD, $\eta \gg |1 - \rho|$,
 the acoustic mode Q is reduced to $\qeff \simeq \eta / \mu$.
When the acoustic mode resonance is well above the AMD resonance,
 $\qeff \simeq \rho / \eta \mu$, and when it is well below the AMD resonance,
 $\qeff \simeq 1 / \eta \mu \rho$, assuming $\eta^2 \ll 1$.

To suppress PIs, the test mass acoustic mode Q-factors only need to be reduced
from $\approx 10^7$ to $10^5-10^6$.
Using this simple model we can estimate that a  $3{\rm g}$ AMD with
 $\eta = 0.1$ on a $10{\rm kg}$ test mass,
 can give $\qeff \lesssim 10^5$ for resonances with $\tfrac{1}{3} < \rho < 3$.

However, this model ignores a number of important factors.
One of these is the location of the AMD relative to the nodes and anti-nodes
 of each test mass acoustic mode.
Quantitatively speaking, the modal mass $M$ of a given mode
 should be increased in this model by the ratio of the displacement at the AMD location
 to that of the mode's anti-node squared $M' = M (x_{\rm max} / x_{\rm AMD})^2$.
Thus, an AMD located near a node will have 
 a reduced value of $\mu$, and will provide little damping.

Other important factors include the multiple coupled degrees of freedom of the AMD
 and the directional nature of the piezo damping material,
 both of which are covered in the follow section.

\begin{table*}[t]
\caption{\label{table:AMDcomp} List of the components for acoustic mode damper prototype. The dimensions and loss angle value are used in finite element modeling described in Sec. \ref{sec:FEM}}. 
\centering 
\begin{ruledtabular}
\begin{tabular}{cccc} 
Component  & Material &  Dimenssions& Loss angle\\
Resistor & Ceramic, surface mount & 0.7$\times$1.2$\times$0.5 mm$^3$ & 10 k$\Omega$\\
Reaction mass & Tungsten &  0.7$\times$1.2$\times$0.5 mm$^3$ & 4e-5\footnotemark[1] \\
Epoxy & conductive TruDuct 2902 & 25 $\mu m$ & 0.15\footnotemark[1]\\
Piezo & shear plate TRS 200HD & 0.7$\times$1.2$\times$0.5 mm$^3$ & 0.014\\
Base & fused silica + gold coating & 0.7$\times$1.2$\times$0.5 mm$^3$ & 7.6$\cdot10^{-12}\cdot$f $ ^{0.77}$\footnotemark[2] \\
Epoxy & non-conductive (EP30, MasterBond) & 25 $\mu$m & 0.1 \footnotemark[1]\\[1ex]
\end{tabular}
\end{ruledtabular}
     \footnotetext[1]{Assumed loss value.}
     \footnotetext[2]{Loss obtained from \cite{Penn20063}.}
\end{table*}

\subsection{Finite Element Model (FEM)}\label{sec:FEM}

The simple 1-D model introduced in the previous section is useful to provide an intuitive understanding of the AMD damping mechanism. However, it is inadequate to represent the details of the interaction between the AMD and the test mass acoustic modes (TMAMs).
Each AMD has at least six resonant modes and hundreds of TMAMs are present in the frequency band of interest; a Finite Element Model is required in order to properly reproduce these modes and calculate the mode overlap between them.

A FEM of the Advanced LIGO test mass with two attached AMDs was constructed with the ANSYS program \cite{ANSYS}.
The AMD model corresponds to the parameters of our prototype AMD (see Table \ref{table:AMDcomp}), and the test mass model parameters are reported in Table \ref{table:TM}.  
All dissipation mechanisms in the test mass substrate, coating and bonds were included in the FEM,
 along with all the losses related to the acoustic mode damper structure
  (see table \ref{table:TNres} for a full list).

A piezoelectric material (PZT), for which energy dissipation can be easily controlled,
 is ideal for AMD construction.
The AMD design modeled here has 2 PZTs sandwiched between a reaction mass
 and the interferometer test mass (see figure \ref{fig:spring-mass}).
The PZTs operate in shear, and are poled orthogonally to ensure that all but 1 of the
 6 lowest frequency AMD modes are damped.
See table \label{tab:AMDreso} for a list of modes; the compression mode is not damped
 by the shear plates.
 
For the shear resonant damper, the spring constant can be associated with the shear deformation
of the piezomaterial
\begin{equation}
k(1+i\eta) = Re(c^{su}_{55})\left(1 + i \eta^{pzt}\right)\frac{S}{h},
\end{equation}
where $c_{55}$ is the shear stiffness matrix element, $S$ is the area, and
$h$ is the height of the shear plate, respectively.\\
The loss factor  $\eta^{pzt}$ is induced by shunting the shear plate with a
resistor. The active stiffness component $c_{55}$ becomes a complex quantity
with a nonzero imaginary stiffness due to the shunt. The
magnitude of $Im(c_{55})$ is strictly related to the impedance of the
shear plate-resistor circuit. The loss factor of the shunted shear plate
can be defined as
\begin{equation}
\eta^{pzt} = \frac{Im(c^{su}_{55})}{Re(c^{su}_{55})}.
\end{equation}

Because the impedance of any PZT is capacitive, the loss factor
$\eta^{pzt}$ is frequency dependent.
As such, careful selection of the shunting resistor and PZT
dimensions can be used to maximize loss in the band of interest,
 as shown in figure \ref{fig:PZTloss}.
A more detailed discussion of the
 piezoelectric loss angle can be found in Appendix \ref{a:pzt}.

We validated the FEM of the AMD by computing its principle resonances,
 and comparing them with direct measurements performed on a prototype AMD.
A total of five principle resonances were identified, with three types of modes: two flag, two anti-flag, and one rotation mode (see Table \ref{tab:AMDreso}). All these modes are characterized by large shear deformation for the double piezo configuration in the AMD. The sixth compression mode was not measured as it does not involve shear of the piezo plate.
Table \ref{tab:AMDreso} shows the good agreement between the output of the model and measurements on the AMD prototype.

\begin{table}[t]
\caption{Verification of the finite element model for AMD. Five principle resonances (two flag, two anti-flag and one rotational mode) obtained with the model are compared to the measured values. The x, y suffix corresponds to the shear plate polarization direction. The principle resonances of the AMD were measured with a capacitive bridge circuit where one of the matching capacitors was the AMD prototype. The difference between flag and anti-flag modes corresponds to the location of the rotation axis about which reaction mass rocks. For the flag pole the rotation axis is at the bottom surface of the shear plate whereas for the anti-flag mode the rotation axis is at the height of mass center of the reaction mass.} 
\centering 
\begin{tabular}{c c c} 
\hline
\hline 
Mode type  & FEM [kHz] &   Measured [kHz]\\[0.5ex]
Fy-mode & 9.96 & 9.77\\
Fx-mode & 12.87 & 12.61\\
R-mode & 23.28 & 24.13\\
aFy-mode & 38.36 & 37.39\\
aFx-mode & 50.51 & 48.86\\[1ex]
\hline 
\end{tabular}
\label{tab:AMDreso} 
\end{table}

Harmonic analysis (finite element analysis with an excitation at a fixed frequency) was conducted to estimate the Q-factor of each of the TMAMs. 
This approach allows us to include frequency dependent variables
 such as the shear plate stiffness $\Re(c_{55}^{su})$, and its loss angle (see Appendix \ref{a:pzt}).

\begin{figure}[h!]
    \center{\includegraphics[width=0.5\textwidth]{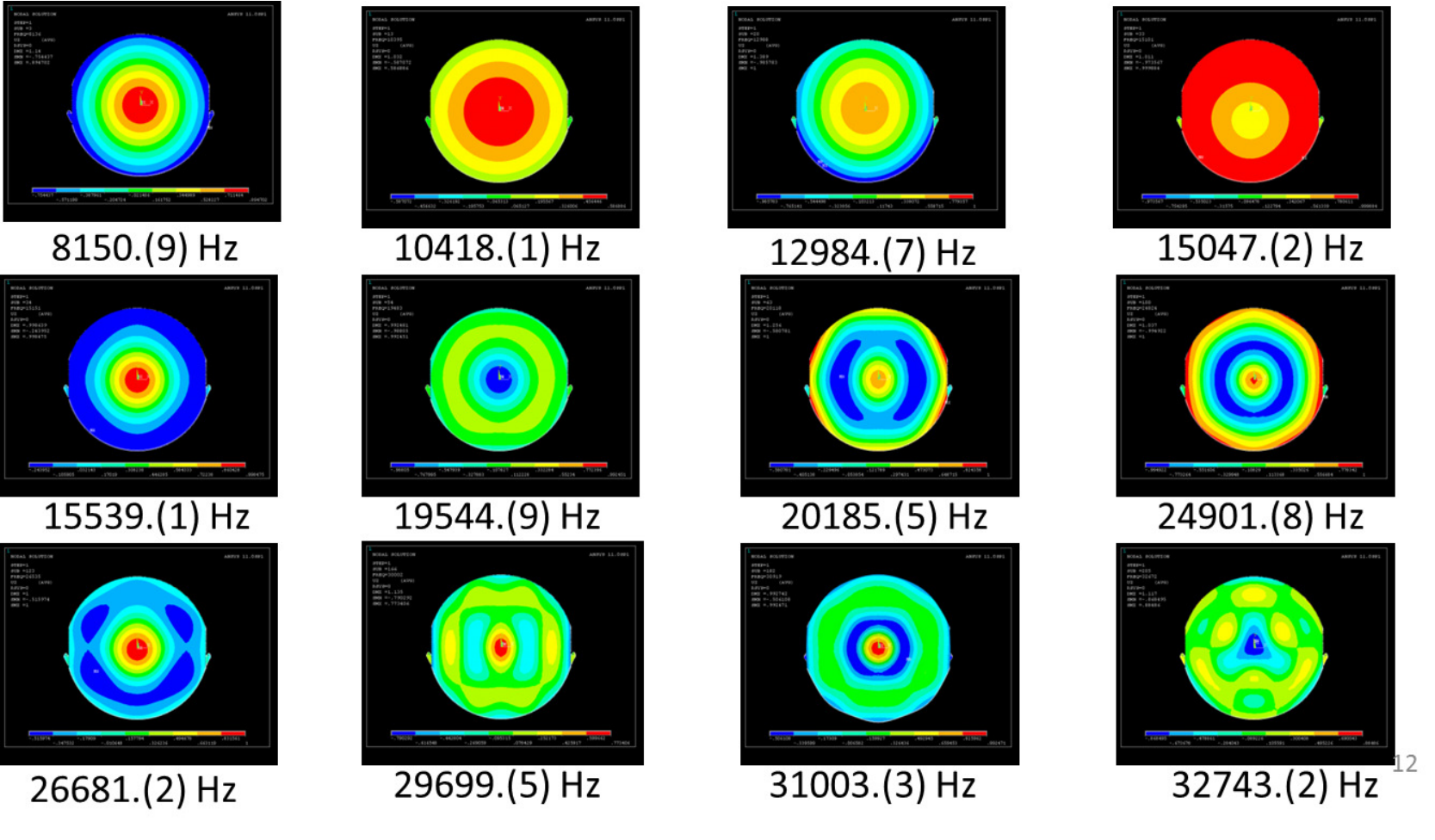}}
      \caption{\label{fig:modes} Test mass drumhead modes for which Q-factor was measured. The color-code corresponds to the test mass front-face displacement amplitude. This figure was obtained from FEM analysis of the test mass model without AMDs.}
\end{figure}

The modal Q-factor of each resonant mode of the system was calculated as
\begin{equation}
Q(f_n) =  \frac{\sum_{m}E_m(f_n)}{\sum_{m} E_m(f_n) \, tan(\phi_{m}) }\label{eq:eff_loss},
\end{equation}
where $f_n$ is the frequency of the $n^{\rm th}$ acoustic mode (see figure \ref{fig:modes}).
The loss associated with each structural component is treated separately;
 $E_m$ is the modulus of the strain energy in the $m^{\rm th}$
 component and $\tan(\phi_{m})$ is its loss factor  (see tables \ref{table:AMDcomp} and  \ref{table:TM}).

The modeled resonant frequencies of 12 TMAMs were then compared to the measured resonance frequencies (see table~\ref{table:TMresults}), obtained according to methods described in section~\ref{sec:er}. The front face displacement of the modes are shown in Fig. \ref{fig:modes}.

Note that the agreement between calculated and measured Q-values increases with frequency.
This may indicate an additional dissipation process which is missing in the model. Nevertheless, a small relative frequency offset $\Delta f$ below 1\% in Table \ref{table:TMresults} indicates a good agreement of the FEM with measured values. 

\begin{table}[h!]
\caption{The Q-factor of the test mass modes without AMDs installed. The 4th and 7th column correspond to the relative frequency noise $\Delta$f\% and Q-factor ratio between calculated and measured value, respectively.}
\centering 
\begin{tabular}{c c c c  c c c} 
\hline
\hline 
Mode \# & Freq. [Hz] & Freq. [Hz]   &  $\Delta$f \%    &   & Q-factor  &   \\[0.5ex]
        & FEM        &  Measured    &                 &     FEM   & Measured  & Ratio               \\[0.5ex]%
\hline
1 & 8128.3 & 8150.9& 0.3 & 37M& 1.9M & 19.5  \\
2 & 10391.1 &10418.1 & 0.3 & 63M& 14M & 4.5\\
3 & 12999.1 &12984.7 & 0.1 & 29M& 15M & 1.9\\
4 & 15101.4 &15047.2 & 0.4 & 56M& 17M & 3.3\\
5 & 15151.0 &15539.1 & 2.5 & 55M& 16M & 3.4\\
6 & 19487.0 &19544.9 & 0.3 & 30M& 7.0M & 4.3\\
7 & 20113.6 &20185.5 & 0.3 & 27M& 13M & 2.1\\
8 & 24824.0 &24901.8 & 0.3 & 32M& 16M & 2.0\\
9 & 26504.3 &26681.2 & 0.7 & 48M& 18M & 2.7\\
10 & 29767.4 &29699.5 & 1.0 & 18M& 15M & 1.1\\
11 & 30912.1 &31003.3 & 0.3 & 18M& 12M & 1.5\\
12 & 32664.1 &32743.2 & 0.2 & 14M & 13M & 1.1 \\[1ex]
\hline 
\end{tabular}
\label{table:TMresults} 
\end{table}

\section{Experimental Results} \label{sec:er}

Several AMD prototypes were constructed, each consisting of six components (see Table \ref{table:AMDcomp}):

{\it Reaction Mass}: A 10 g tungsten cylinder which tunes the AMD
 principle resonances to frequencies above 10 kHz is located on top of the shear plates.

{\it Shear Plate}: Two piezoelectric shear plates, oriented with perpendicular polarizations
 are glued to the reaction mass and base with conductive epoxy Tra-Duct2902.
The epoxy serves to electrically connect the PZT electrodes to the reaction mass and base.

{\it Base}: The interface between flat shear plates and curved barrel of the test mass. The top flat surface is gold coated with separate sections to which shear plates are glued. The bottom surface is curved and match to the test mass barrel curvature. The base is made from fused silica and glued to the test mass with nonconductive epoxy.

{\it Shunting circuit}: One 10 k$\Omega$ resistor for each shear plate is glued to the reaction mass with conductive epoxy. The circuit is closed with 100 $\mu$m diameter silver coated copper wire,
 which is  soldered to the resistor and gold coated surface of the base.\\ 

To measure Q-factors of the test mass modes, both with and without AMDs attached, a 16 m optical cavity was used, with the test mass forming the end mirror of the cavity. The optical cavity supports a 2 mm diameter Gaussian mode; the resonant beam probes the motion in the center of the test mass face, so that modes with an anti-node in the center of the mass are easily measured. A total of 12 modes were identified (Fig.~\ref{fig:modes}) and measured, with the results shown in Table~\ref{table:TMresults}.

In order to measure the impact of the AMD on the Q of the TMAMs,
 several modes of the test mass were excited using  electrostatic actuators \cite{0264-9381-29-11-115005}
 and their ring-down times observed.
This measurement was repeated in three configurations:
 with no AMD, with 2 AMDs, with 2 AMDs which lacked resistive damping (shunt wires cut).

Each acoustic mode was detected in the cavity locking error signal as a peak in the Fourier domain.
After being excited, each mode amplitude was recorded as a function of time to estimate the
 decay time $\tau$.
 The Q-factor was determined from the decay time $\tau$, according to
\begin{equation}
Q =  \pi f_{0} \tau \label{eq:q} .
\end{equation}
where $f_{0}$ is the resonant frequency in Hz and $\tau$ is the exponential decay time constant. 

    \begin{figure}[ht]
    \center{\includegraphics[width=0.5\textwidth]{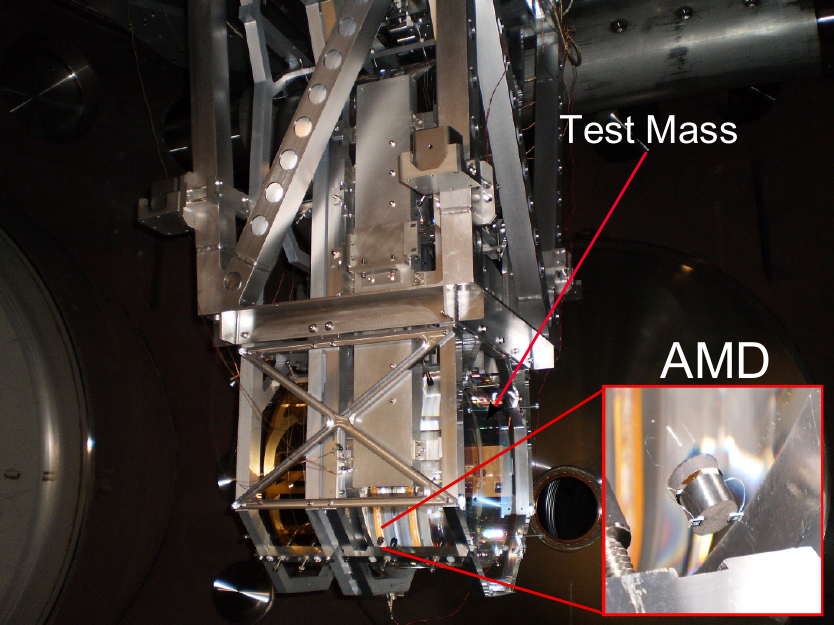}}
      \caption{\label{fig:amdexp} Two acoustic mode dampers (AMDs) attached to
       the barrel of a suspended aLIGO test mass.
       The AMDs are on opposite sides of the test mass, $\sim45^\circ$ down from the midline.}
    \end{figure}
    
Measurements of the test mass mode Q-factor performed after installing two AMDs clearly show the substantial damping capability of the AMD prototype, as reported in Table \ref{table:TMresults} (see also Table \ref{table:AMDresults} in Appendix for additional details).

The results indicate that out of 12 modes, 11 are suppressed by at least factor of 10
 and in some cases by more than two orders of magnitude.
The relatively large discrepancy between model and measurement for mode
 \#2 can be explained by the AMD and TM interaction condition for this particular mode.
The FEM predicts that mode \#2 will be within 500Hz of the AMD Fy-mode at 10 kHz, while the measured values give a separation of  800Hz (mostly due to the AMD resonance being off), reducing the interaction between the AMD and TM mode.

In the off-resonance interactions,
 which are more common and set the lower limit to AMD performance, 
 the discrepancy is generally less than a factor of a few.

Surprisingly, modes \#2 and \#6 show a counterintuitive behavior; 
 opening the resistive circuit of the AMDs to decreases the TMAM Q-factor.
However, since electrical circuit of AMD affects mechanical stiffness of the shear plates
 it is expected that principle resonances also changed when the circuit is opened.
Also, if the AMD resonance is close to the TMAM frequency, equation \ref{eq:qeff}
 indicates that the mode Q can be decreased by \emph{lowering} the AMD loss.

The large Q reduction for modes \#3 and \#4 is due the
 on-resonance interaction between TMAM and AMD,
 whereas for modes \#5 and \#12 the large Q reduction
 is due to the anti-node AMD location on the TMAM.

\begin{table}[ht]
\caption{Test mass mode suppression obtained with two AMDs.} 
\centering 
\begin{tabular}{c c c c} 
\hline
\hline 
Mode \# & Freq. [Hz] & Damping & Resistive \\[0.5ex]
        &            &  factor        & contribution [\%]  \\ 
\hline
1 & 8150.9   & 32.2   &    12 \\
2 & 10418.1  & 31.8   &   -57\\
3 & 12984.7  & 441.2  &    21\\
4 & 15047.2  & 81.0   &    9\\
5 & 15539.1  & $>$320 & $>$71\\
6 & 19544.9  & 15.6   &    -2\\
7 & 20185.5  & 13.8   &    41\\
8 & 24901.8  & 5.2    &     4\\
9 & 26681.2  & $>$360 &  $>$0\\
10 & 29699.5 & 23.8   &    43\\
11 & 31003.3 & 307.7  &    68\\
12 & 32743.2 & $>$260 & $>$4 \\[1ex]
\hline 
\end{tabular}
\label{table:results} 
\end{table}

\section{Analysis of the AMD thermal noise} \label{sec:tn}

The AMD is designed to increase the mechanical damping of the test mass acoustic modes
above 10 kHz. At the same time, the AMD must introduce minimal additional mechanical 
loss in the 10-1000 Hz band, where low mechanical loss is required to keep test mass
thermal noise small \cite{FDT}. Thus it is critical to calculate the thermal noise
resulting from the AMDs in our overall evaluation of this PI mitigation technique.

We used the FEM described above to calculate the thermal noise resulting from our 
experimental test of two AMDs attached to a test mass.
The AMD thermal noise was calculated numerically at 100 Hz,
 the most sensitive part of the detection band, using Levin's approach \cite{PhysRevD.57.659}.
The energy dissipation per cycle was computed using Eqn. \ref{eq:losseff}
 for a pressure profile corresponding to the Advanced LIGO geometry
 (laser beam radius $\omega_{0} = 5.5$ cm, incident on the front face of the TM).
Results are shown in Table~\ref{table:TNres}.

The test mass thermal noise level of $5.2 \times 10^{-21}{\rm m/\sqrt{Hz}}$ corresponds
to the design level for Advanced LIGO, and is dominated by optical coating loss~\cite{0264-9381-19-5-305,Evans2008}.
As Table~\ref{table:TNres} shows, while our \emph{prototype} AMD would contribute
 significantly to thermal noise it is not orders of magnitude above the more fundamental
 sources of thermal noise.

\begin{table}[ht]
\caption{Thermal noise budget for test mass and acoustic mode damper prototype when attached to the test mass. The thermal noise was calculated for the laser beam spot size of $\omega$ = 5.5 cm.} 
\centering 
\begin{tabular}{l c} 
\hline
\hline 
Component & Thermal noise @ 100 Hz \\
 &  $[10^{-21} m / \sqrt{Hz}]$\\[0.5ex]
\hline
 \multicolumn{2}{l}{\bf{Test Mass:}}\\
Substrate & 0.8 \\ 
Optical coating & 5.1\\
Suspension ears & 0.0 \\ 
Ears bond & 0.6\\
Total per TM & {\bf{5.2}}\\
 \multicolumn{2}{l}{\bf{AMD:}}\\
Reaction mass (RM)& 0.3\\ 
Epoxy (RM-PZT)& 4.8\\
Shear plate (PZT-X)\footnotemark[1] & 5.5\\
Shear plate (PZT-Y)\footnotemark[1] & 5.5\\
Epoxy Base-PZT & 7.2\\
Base & 0.0 \\ 
Epoxy Base-TM & 6.0\\
Total per AMD & {\bf{13}}\\
\hline 
\end{tabular}
\label{table:TNres} 
\footnotetext[1]{A product of a structural+resistive loss angle. Note that a major contribution to that value comes from the structural loss, see Fig.~\ref{fig:PZTloss}.}
\end{table}

Relative to the prototype device,
our model points to several design and material improvements that can be made
to significantly reduce the thermal noise impact. The major AMD thermal noise 
contributors are the epoxies used to bond the AMD elements, and structural
loss in the piezo shear plates. The former can be improved with lower loss
epoxy and thinner bond layers. The latter can be improved with a more suitable
choice of piezo material.

Other design elements can also be altered. The mass of the reaction mass can be
reduced to lower the thermal noise without significantly affecting the acoustic
mode damping performance. Another modification would be to avoid alignment of the 
piezo shear plate polarization with the laser beam axis, to minimize the contribution 
of resistive loss to the thermal noise. These and other design optimizations will
be explored in a future paper.


\section{Conclusion} \label{sec:con} 

Acoustic mode dampers represent a simple yet effective approach to damping parametric instabilities.
The great advantage of this approach over active damping \cite{Miller2011788} is that
 many test mass acoustic modes are effected simultaneously, and no further intervention
 is necessary.
This is likely to be a critical feature in instruments that would otherwise suffer from multiple
 acoustic modes simultaneously excited by parametric instabilities.

The investigation presented here involved modeling and construction of a prototype AMD,
 which was shown to effectively damp test mass acoustic modes.
The thermal noise associated with this prototype AMD was also computed.
Though the prototype AMD does not meet
 the stringent thermal noise requirements of gravitational wave detectors,
 several design elements were identified for improvement,
 making this approach a viable solution to parametric instabilities.

\begin{acknowledgments}
The authors gratefully acknowledge the support of the National Science
Foundation and the LIGO Laboratory, operating under cooperative Agreement No. PHY-0757058
This paper has been assigned LIGO Document No. LIGO-P1400257.
\end{acknowledgments}

\bibliography{AMDbib}

\appendix

\def\zhat{\widehat{Z}}

\section{\label{a:pzt} Piezoelectric material}

Any piezoelectric material is strictly characterized by electromechanical
properties. The fact that both electrical and mechanical properties are
inseparable allows us to represent the dissipation process in a shunted
piezoelectric material either as a Jonson heat or mechanical loss. For the
purpose of this work we focus on disspation process in terms of the
mechanical loss.\\
From the stress-charge of the piezoelectric constitutive equation it is
straightforward to derive the total induced stress in the shunted
piezoelectric material  \cite{Hagood1991243}. Assuming no external current
plugged to the piezoelement electrode we get
\begin{equation}
\boldsymbol{\sigma}_{(6\times 1)} = \left(\boldsymbol{c}^{E}_{(6\times
6)}-i\omega\frac{\boldsymbol{e}^{t}_{(6\times
3)}\boldsymbol{Z}^{TOT}_{(3\times 3)}\boldsymbol{A}_{(3\times
3)}\boldsymbol{e}_{(3\times 6)}}{\boldsymbol{L}_{(3\times
3)}}\right)\boldsymbol{S}_{(6\times 1)}  \label{eqA:stress},
\end{equation}
where $\boldsymbol{c}^{E}$ is the mechanical stiffness matrix under
constant electric field, $\boldsymbol{e}^{t}$ and $\boldsymbol{e}$ are the
piezoelectric stress constant. The upper script $t$ refers to transpose
operator \cite{26560}. Electrode area is represented by matrix
$\boldsymbol{A}$, whereas thickness of the piezomaterial between
electrodes by matrix $\boldsymbol{L}$. The strain $\boldsymbol{S}$ is a
product of the acting stress $\boldsymbol{\sigma}$ on piezo-element and
charge accumulation in the piezo-element. Each bracket corresponds to the
matrix dimenssion. It is assumed that piezoelectric element has a brick
shape thus matrices $\boldsymbol{Z}^{TOT}$,$\boldsymbol{A}$, and
$\boldsymbol{L}$ are diagonal.\\
The total impedance in Eqn.\ref{eqA:stress} is inverse sum of the
piezoelement admittance under constant electric field $\boldsymbol{Y}^E$
and the admittance of the external circuit  $\boldsymbol{Y}^{SU}$
connected to the piezoelement electrodes, thus
\begin{equation}
\boldsymbol{Z^{TOT}} =
\left(\boldsymbol{Y}^E+\boldsymbol{Y}^{SU}\right)^{-1} 
\label{eqA:impedance}.
\end{equation}
The admittance of piezelement $\boldsymbol{Y}^E$ is assumed to be
exclusively capacitive. Hence,
\begin{equation}
\boldsymbol{Y}^E =
i\omega\boldsymbol{A}\boldsymbol{\epsilon}^S\boldsymbol{L}^{-1}=i\omega\boldsymbol{C}^S=i\omega\boldsymbol{C}^T\left(\boldsymbol{\epsilon}^T\right)^{-1}\boldsymbol{\epsilon}^S
 \label{eqA:piezoadmitance},
\end{equation}
where  $\boldsymbol{C}$,  $\boldsymbol{\epsilon}$  is the capacitance and
the dielectric constant matrices under constant strain $S$ and constant
stress $T$, respectively. It is convenient to operate with
$\boldsymbol{C}^T$ since this quantity can be easily measured at stress
free conditions and no shortened piezelement electrodes.\\
According to Eqn.\ref{eqA:stress}, if the piezoelement is integrated with
a nonzero impedance electric circuit, the imaginary part stiffness tensor
arises. The imaginary part can be interpreted as a dissipative component
of the stiffness tensor. The total shunted stiffness matrix is
\begin{equation}
\boldsymbol{c}^{SU} =
\boldsymbol{c}^{E}-i\omega\boldsymbol{e}^{t}\boldsymbol{Z}^{TOT}\boldsymbol{A}\boldsymbol{e}\boldsymbol{L}^{-1}
\label{eqA:comp} .
\end{equation}
Because matrices $\boldsymbol{A}$, $\boldsymbol{Z}^{TOT}$, and
$\boldsymbol{L}$ are diagonal and knowing that
$\boldsymbol{A}\boldsymbol{L}^{-1} =
\boldsymbol{C}^{T}(\boldsymbol{\epsilon}^T)^{-1}$ Eqn. \ref{eqA:comp} can
be written as
\begin{equation}
\boldsymbol{c}^{SU} =
\boldsymbol{c}^{E}-\boldsymbol{e}^{t}\boldsymbol{e}\boldsymbol{\zhat}\left(\boldsymbol{\epsilon}^S\right)^{-1}
\label{eq:comp_final},
\end{equation}
where $\boldsymbol{\widehat{Z}}$ is the nondemensional impedance
\begin{equation}
\boldsymbol{\zhat} = \boldsymbol{Z}^{TOT}\boldsymbol{Y}^E \label{eqA:imp}.
\end{equation}
Note, for zero impedance shunted circuit sets $\boldsymbol{\zhat}$ to be a
unity matrix $\boldsymbol{\zhat}=\boldsymbol{I}$ whereas for the nonzero
impedance shunting $\boldsymbol{\zhat}$ is a complex quantity. Using
indexing  Eqn.\ref{eqA:comp} can be written as
\begin{equation}
c_{ij}^{su} = c_{ij}^{E}- \frac{e_{ki} e_{kj}\zhat_{kk}}{\epsilon_{kk}^S} 
\label{eqA:general} .
\end{equation}
where $k = 1,2,3$ and corresponds to the electrode position,
$\boldsymbol{\epsilon}^S$ is the dielectric permittivity of the
piezoelectric material.

From Eqn.\ref{eqA:piezoadmitance} and \ref{eqA:general} becomes clear that
loss can be control with the shunting circuit. Moreover, such loss is in
fact a frequency dependent quantity with arbitrary shape of the loss curve
dependent on shunting circuit.\\

For the shear plate with the single pair of electrodes only a stiffness
matrix element $c_{55}^{su}$ is affected by shunting resistor, thus
\begin{equation}
c_{55}^{su} = c_{55}^{E}- \frac{e_{15} e_{15}\zhat_{11}}{\epsilon_{11}^S} 
\label{eqn:55} .
\end{equation}
and the nondemensional complex impedance $\zhat_{kk}$ has a single
nonunity element
\begin{eqnarray}
\zhat_{11} =  \frac{i\omega\epsilon_{11}^S RC^T}{i\omega\epsilon_{11}^S
RC^T+\epsilon_{11}^T} = \nonumber\\
\frac{\left(\omega\epsilon_{11}^S
RC^T\right)^2}{\left(\omega\epsilon_{11}^S
RC^T\right)^2+\left(\epsilon_{11}^T\right)^2}+\nonumber\\
i\frac{\omega\epsilon_{11}^S\epsilon_{11}^T
RC^T}{\left(\omega\epsilon_{11}^S
RC^T\right)^2+\left(\epsilon_{11}^T\right)^2}  \label{eqn:z11}.
\end{eqnarray}
Since the stiffness matrix $\bm{c}^{su}$ is a complex quantity we can
define the loss factor $\bm{\eta}$ as
\begin{equation}
\bm{\eta} = \frac{Im(\boldsymbol{c}^{su})}{Re(\boldsymbol{c}^{su}) }
\label{eqA:loss1},
\end{equation}
and thus
\begin{equation}
\eta_{kij} =  \frac{Im(\zhat_{kk})\chi_{kij}}{Re(\zhat_{kk})\chi_{kij}+1}
\label{eq:loss2},
\end{equation}
where $\chi_{kij}= e_{ki} e_{kj} (c_{ij}^{E}\epsilon_{kk}^S)^{-1}$. Index
$k$ is associated with the electrodes orientation, whereas indices $i$ and
$j$ correspond to stress-strain directions in the stiffness matrix
$\boldsymbol{c}^{su}$.
The shunting loss factor of the shear plate becomes
\begin{equation}
\eta_{155} = \frac{\Im(\zhat_{11})\chi_{155}}{\Re(\zhat_{11})\chi_{155}+1}\label{eq:loss3}.
\end{equation}
This is the main loss mechanism based on which shear AMD operates. Note
that according to Eqn.\ref{eqn:55}, the material stiffness (the real part
of $c_{55}^{su}$) is also reduced and should be included in analysis.\\

\subsection{Total mechanical loss angle}
An assumption of the stiffness matrix being real in the absence of shunt
is not sufficient for accurate estimation of energy dissipation. It
becomes especially important in the thermal noise analyisis, see Sec.
\ref{sec:tn}.\\ For known material loss angle of the nonshunted
piezoelement, the total loss factor of piezoelectric material can be
defined as
\begin{equation}
\eta_{kij}^{tot} = 
\frac{tan(\phi_{ij}^{mat})+Im(\zhat_{ii})\chi_{kij}}{1+Re(\zhat_{ii})\chi_{kij}}
\label{eqA:losstot},
\end{equation}
where $\tan(\phi_{ij}^{mat})$ is the material loss factor  matrix of
piezo-element. The loss factor $\tan(\phi_{ij}^{mat})$ of
$\boldsymbol{{c}^E}$ can be easily computed using Eqn. \ref{eqA:loss1}.

All piezoelectric material has anisotropic structure what implies that
strain energy dissipation in such material must depend on the piezoelement
geometric shape. It is more convenient to use effective loss angle
$\phi_{eff}$ rather than the loss angle $\phi$ for each stiffness matrix
component. We define the energy dissipation per cycle in the piezo-element
\begin{eqnarray}
W_{dis} &=& 2 \pi W_{st} \eta_{eff} \\
 &=& \int{Re(S_{i}S_{j}^{*}) Re(c_{ji}^{su})\eta_{ji}^{tot}}dV \label{eq:losseff}
\end{eqnarray}
where $W_{st}$ is the stored strain energy such that $2\pi W_{st} =
\int{Re(S_{i}T_{j}^{*})}dV$, where S, T are the complex strain and stress,
respectively and V is the volume of the piezo-element.\\
Since loss factor of the piezoelement is inverse of its Q-factor or
equally a ratio of dissipated energy per cycle $W_{dis}$ to energy stored 
$W_{st}$ in the piezoelement, we can defined  the effective noise fuctor
$\eta_{eff}$
\begin{equation}
\eta_{eff} =  \frac{W_{dis}}{2\pi W_{st}}=\frac{\int Re(S_{i}S_{j}^*)
Re(c_{ji}^{su})\eta_{ji}^{tot}dV}{\int Re(S_{i}T_{j}^{*})dV}
\label{eqA:losseff},
\end{equation}
where $S$, $T$ are the complex strain and stress, respectively and $V$ is
the volume of the piezo-element. This is a key equation in the finite
element analysis which properly estimate contribution of piezoelemnt in
the strain energy dissipation in the test mass.

In our analysis we assumed constant material intrinsic loss factor for all
$\boldsymbol{c}^E$ elements equal to $\tan(\phi_{ij}^{mat})$=0.014 
\footnote{The information about the imaginary part of the stiffness matrix
is not available. We assumed that the material loss factor corresponds to
the Q-factor of the piezomatrial provided by the manufacturer. In our
opinion low Q PZT should have fairly constant loss for all stiffness
components}, what leads to
\begin{equation}
\eta_{155}^{tot} = 
\frac{0.014+Im(\zhat_{11})\chi_{155}}{1+Re(\zhat_{11})\chi_{155}}
\label{eq:losstot} .
\end{equation}
Note, the remaining elements of shunting induced loss $\eta_{kij}^{tot}$
are equal to the material structural loss factor. For the shear plate
configuration $\chi_{155}$ is simply a function of electromechanical
coupling coefficient $k_{15}$ and is equal to $\chi_{155} =
\frac{k_{15}^2}{1-k_{15}^2}$. The total loss angle for the $c_{55}^E$
stiffness element is shown in Fig.\ref{fig:PZTloss}.

\section{Additional Tables}
\begin{table*}[h]
\caption{\label{table:TM}Values for the aLIGO end test mass parameters used in this paper.}
\footnotesize\rm
\begin{ruledtabular}
\begin{tabular}{lll}
{\it\bf Optical parameters:}&&\\
End test mass transmissivity  &5.0 ppm&\cr
\multicolumn{3}{l}{1 layer each of Ta$_2$O$_5$/SiO$_2$ with thickness of 2.32 $\mu$m, and 3.49 $\mu$m\footnotemark[1], respectively}\\

\hline
\multicolumn{3}{l}{\it \bf Mechanical properties of optical coating and ear bond:}\\
&Ta$_2$O$_5$&SiO$_2$\hspace*{2.6cm} Ear bond\\
Young's modulus&140 GPa&70 GPa\hspace*{2.2cm} 7.2 GPa\\    
Poisson ratio&0.23&0.17\hspace*{2.9cm}0.17\\
Density&8300 kg/m$^3$&2201 kg/m$^3$\hspace*{1.7cm}2202 kg/m$^3$\\
Refractive  index&2.06539&1.45\hspace*{2.9cm}-\\
Loss angle\footnotemark[2] &2.4$\cdot10^{-4}$+f$\cdot$1.8$\cdot10^{-9}$ &0.4$\cdot10^{-4}$+f$\cdot$1.1$\cdot10^{-9}$\hspace*{0.5cm}0.1\\
&&\\
\hline
{\it \bf Test mass dimensions:}&&\\                                                 
Radius&0.17  m&\\                                           
Thickness&0.2  m&\\  
Flats width&0.095 m&\\                                       
Wedge angle &0.07  deg&\\                                       
Mass&40  kg&\\
Loss angle\footnotemark[3] &7.6$\cdot10^{-12}\cdot$f $ ^{0.77}$ &\\ 
Material\footnotemark[3] & fused silica\\ 
\end{tabular}\\
 \footnotetext[1]{For the purpose of numerical analysis, a multilayer coating was reduced to double layers with a total thickness which corresponds to the sum of all 18/19 layers of Ta$_2$O$_5$/SiO$_2$ for the ETM.} 
 \footnotetext[2] {f-frequency. Note, the difference in loss angles for the substrate, and for the optical coating made from fused silica. Losses were obtained from \cite{0264-9381-24-2-008}. Additionally, coating loss angles were revised to the current measured values.}
 \footnotetext[3] {Both test mass and suspension ears.}
\end{ruledtabular}
\end{table*}

\begin{table}[h!]
\caption{Results of the Q-factor measurement with attached AMDs for shunted and non-shunted cases.} 
\centering 
\begin{tabular}{c c c c c} 
\hline
        & \multicolumn{4}{c}{Q-factor} \\[0.5ex]
Mode \# & \multicolumn{2}{c}{Resistive} & \multicolumn{2}{c}{Open Circuit}\\
        & FEM      & Measured  & FEM & Measured \\
\hline
1  & 52k & 59k    & 70k & 67k   \\
2  & 7.9k & 440k    & 9.9k & 280k   \\
3  & 23k & 34k    & 23k & 43k   \\
4  & 420k & 210k    & 510k & 230k   \\
5  & 2.4M & $<$50k & 2.2M & 170k   \\
6  & 1.9M & 450k    & 4.8M & 440k   \\
7  & 1.1M & 940k    & 1.6M    & 1.6M    \\
8  & 6.7M & 3.1M    & 9.5M & 3.2M   \\
9  & 49k & $<$50k & 64k & $<$50k \\
10 & 1.9M & 630k   & 3.3M & 1.1M   \\
11 & 61k & 39k      & 116k    & 120k    \\
12 & 25k & $<$50k & 260k & 52k   \\[1ex]
\hline 
\end{tabular}
\label{table:AMDresults} 
\end{table}

\end{document}